\documentclass[12pt,preprint]{aastex}

\def\cgcg{CGCG~049-033\,}

\shorttitle{A giant radio jet in \cgcg}
\shortauthors{J. Bagchi et al.}

\received{2007 July 5}
\begin{document}

\title{A giant radio jet ejected by an ultramassive black hole in a 
single-lobed radio galaxy}

\author{Joydeep Bagchi\altaffilmark{1}, Gopal-Krishna\altaffilmark{2},
    Marita Krause\altaffilmark{3} and Santosh Joshi\altaffilmark{1,4}}
\altaffiltext{1}{The Inter-University Centre for Astronomy and 
Astrophysics (IUCAA),  Pune University Campus, Post Bag 4, Pune  411007, 
India; joydeep@iucaa.ernet.in}
\altaffiltext{2}{National Centre for Radio Astrophysics, TIFR, Pune University
Campus, Post Bag 3, Pune  411007, India
}
\altaffiltext{3}{Max-Planck-Institut f\"{u}r Radioastronomie, Auf dem H\"{u}gel 69, 53121 Bonn, Germany
}
\altaffiltext{4}{Aryabhatta Research Institute of Observational Sciences
(ARIES), Manora Peak, NainiTal  263129, India}


\begin{abstract}
We report the discovery of a very unusual, highly asymmetric radio galaxy
whose radio jet, the largest yet detected,
emits strongly polarized synchrotron radiation and can be
traced all the way from the galactic nucleus to the hot spot located $\sim 440$
kpc away.  This  jet  emanates from an extremely
massive black-hole $( > 10^9 M_{\odot}$) and forms a strikingly compact
radio lobe. 
No radio lobe
is detected on the side of the counter-jet,
even though it is similar to the main jet in brightness up to a
scale of tens of kilo parsecs.
Thus, contrary to the nearly universal trend, the
brightness asymmetry in this radio galaxy
{\it increases} with distance from the nucleus. With several unusual
properties, including a predominantly toroidal magnetic field, this
Fanaroff-Riley type II mega-jet is an exceptionally useful laboratory for
testing the role of magnetic field in jet stabilization and
radio lobe formation.
\end{abstract}



\section{Introduction}

Radio sources associated with  Abell clusters of galaxies are usually
weak radio emitters but present a great diversity of morphologies.
During a search for radio emission from rich clusters we noticed a radio 
source towards Abell~2040, which is associated with a bright  B$\sim$14.5 
elliptical galaxy \cgcg\, offset from the cluster center by 
$\sim$22 arcmin (1.1 Mpc, taking $H_{0}=$  71~km s$^{-1}$ Mpc$^{-1}$, 
$\Omega_{M}=0.27$, $\Omega_{\Lambda}=0.73$). The 
redshift $z$ = 0.04464 of this galaxy is close to $z$ = 0.04564 of the
(radio-quiet) central galaxy (UGC09767)
of the cluster \citep{wegner1999}. In the 
VLA NVSS database \citep{condon1998} we noticed that \cgcg coincides with 
a predominantly one-sided radio source, elongated on 10\arcmin\, scale 
($\sim 0.5$ Mpc), but its structural details remained poorly defined due to 
the modest resolution (45\arcsec). 
In view of its potentially huge size we have imaged the radio source, 
as well as taken its optical spectrum. 
Our observations have revealed the source to be  highly unusual 
in terms of radio size, morphology, and polarization, and in being powered 
by an extremely massive black hole.

\section{Observations}

\subsection{GMRT radio observations (aperture synthesis)}

The GMRT \footnote {The GMRT is a national facility operated 
by the National Centre for Radio
Astrophysics of the TIFR, India}
observations were made at 1.28 GHz ($\lambda$ = 23.4~cm) 
using the  128 channel correlator of total bandwidth 
16 MHz \citep{Swarup1991}. A 
sequence of ~ 30-minute scans of the source was taken 
on 2004 July 15,  interspersed 
with the phase calibrator 1445+099. The telescope gain and the bandpass 
were calibrated  using  3C~286.
After excising  bad data and RFI, the calibrated data were 
transformed into the image plane using the standard routines in  AIPS, 
including a few cycles of self-calibration. By applying suitable tapers 
to the visibilities, two maps with synthesized Gaussian beams of FWHM 
3\arcsec\, and 11\arcsec\, were obtained (Fig.~1). The 11\arcsec\, resolution
map (dynamic range $\sim$ 500) shows a pair of well aligned, 
collimated jets emanating from an  unresolved core (92~mJy, $< 3$\arcsec) and 
extending over 675\arcsec\, (= 578 kpc).  
Only the longer southern jet can be traced up to its termination where it
forms a hotspot and a  remarkably compact radio lobe.  Neither the terminal 
point of the (northern) counter-jet, nor its lobe are detected, implying a 
flux density ratio of at least 20
between the southern lobe and any morphologically similar northern lobe.
Details of the inner  2\arcmin\, region can be seen 
on the 3\arcsec\, resolution GMRT map (Fig.~1). 


\subsection{Effelsberg radio observations (single-dish) }

The source was mapped at 8.35 GHz ($\lambda$ = 3.6~cm) with the 100-m
Effelsberg  telescope in 2004 May and July,  
           \footnote{Based on observations with the 100-m telescope of the MPIfR
	             (Max-Planck-Institut f\"{u}r Radioastronomie) at Effelsberg on
		               behalf of the Max-Planck-Gesellschaft},
yielding both total power and polarimetric images (Fig.~2). The 
single-beam receiver 
has two channels with total-power amplifiers and an IF polarimeter, at 
a bandwidth of 1.1 GHz. For pointing, focusing, and flux calibration, 3C~286 
was used. A total of  31 coverages of the field were taken 
by  scanning along and perpendicular to the  jet. The standard
data reduction procedure \citep{Emerson1988} has led to a  
20\arcmin\,$\times$10\arcmin\, map, with a Gaussian beam of FWHM 84\arcsec\, 
and an rms noise of $\sim$260 $\mu$Jy/beam in total power and 
$\sim$100 $\mu$Jy/beam in linear polarization. Despite the large 
beam averaging, an extraordinarily high linear polarization ($p$ 
 = 20 - 50$\%$, average $30\%$) is found along each jet (Fig.~2).

Further, combining our Effelsberg data with the 1.4 GHz NVSS 
linear polarization
map \citep[Fig.~2;][]{condon1998}, we obtained the Faraday 
rotation-measure (RM) which was found to vary smoothly in the range  
$\rm \sim +10$ to $+40$ rad/$\rm {m^{2}}$. These RMs were used to correct 
the observed polarization vectors at 8.35 GHz and thus derive (after a 
$90^{\circ}$ rotation) the  projected magnetic field vectors. 
The magnetic fields of both jets are predominantly transverse to
their axes (Fig.~2). Note that these vectors should be intrinsic 
to the source because foreground RM is negligible, as estimated from the 
known RMs of eight background sources projected near CGCG~0049-033 \citep{S-N1981}.

Remarkably, even in this single dish observation - more sensitive to  
diffuse emission - no trace of a northern lobe is found, down to a 
sensitivity limit of $\sim$1 mJy/beam. 
Nonetheless, the counter-jet is 
clearly detected, albeit only marginally resolved with the 84\arcsec\, beam.
\subsection{Optical spectroscopy}

Since an existing optical spectrum of CGCG~0049-033, taken with the 
2.4-m Hiltner telescope at Kitt-Peak, did not cover the wavelength of 
any prominent emission line [Fig.~2, M. Colles 2007,  private communication],
we  used the newly
commissioned 2-meter telescope of IUCAA, equipped with the IUCAA Faint Object 
Spectrograph and Camera (IFOSC), together with grism~7 (600 grooves/mm), 
to cover a wider spectral range 3800-6840 \AA. A single 15-minute 
exposure was taken through a 1.5\arcsec\, wide  
slit  centered on the galaxy. For data  reduction  the
IRAF package and for wavelength calibration
the  He-Ne arc lamp standard were used.
The reduced spectrum (spectral dispersion  
1.4 \AA/pixel) confirms the strong absorption lines of Mg~Ib and Na~D 
of the Kitt-Peak spectrum, and also shows the
H$_{\beta}$, Ca+Fe E-band and  the TiO absorption features (Fig. 2). However, 
the emission line [OIII] 5007\AA\, remained undetected 
and thus still provides no hint of  AGN activity. A Gaussian-fit to 
the Na~D (5893 \AA) line gives a very large  stellar velocity 
dispersion $\sigma_{\ast} = 375 \pm 35$ km s$^{-1}$, after correcting for the
instrumental broadening. This is consistent with the Kitt-Peak value 
$\sigma_{\ast} = 353 \pm 60$ km s$^{-1}$ \citep{wegner1999}.

\section{Results}
Data on all 660 extragalactic radio jets known until
2000 December \citep{Liu_Zhang_2002} and the subsequent literature reveals that
the main jet of CGCG~0049-033, with a projected length $\sim 440$ kpc,
is the largest known astrophysical jet, seen stretching all the way from the
nucleus to its terminal hotspot. In physical (projected) extent, it is
$\sim 7$ times the bright jet of the unique one-sided quasar 3C~273
\citep{Conway_Davis1993} and nearly 2 times longer than the  jet
of the giant radio galaxy NGC~6251 \citep{Willis1982}. From Figure 1,
despite the rather limited resolution transverse to the jet, there is a
strong hint of its helical morphology. Also evident is the jet's wiggle
on the scale of $\sim 300$ kpc after the initial $\sim$ 30\arcsec\,
($\sim 25$ kpc) length,
up to which point both the jet and
counter-jet are probably well within the ISM of the host galaxy and both
remain straight, well aligned, and similar in apparent brightness
(typical contrast $\sim 1.3$).
But whereas the main jet remains
collimated and forms a lobe, the counter-jet undergoes
an abrupt fading as indicated by the
non-detection of its radio lobe (Fig.~1 \& 2).

Integrated flux densities of the source are 208 $\pm$ 10, 200 $\pm$
15 and 104 $\pm$ 10 mJy at 1.28 GHz (GMRT), 1.4 GHz (NVSS) and 8.35 GHz
(Effelsberg), respectively.
The spectral index values ($\alpha $; S$_{\nu}$$\propto \nu^{-\alpha}$)
are  0.23, 0.50, 0.58, and 0.37 for the core, jet, lobe, and the
integrated emission respectively, obtained by comparing the NVSS map (1.4 GHz,
smoothed to 84\arcsec\,) and Effelsberg map (8.35 GHz) -- the two 
being most sensitive to extended emission.
Note that the integrated luminosity 
($8.8 \times 10^{30}$ \, ergs~s$^{-1}$~Hz$^{-1}$ at 1.4 GHz) falls 
nearly 2 orders-of-magnitude below the
Fanaroff-Riley transition \citep{FR74}, making the FR II morphology
of this jet  very unusual. However, an FR I pattern 
characterized by  jet symmetry  is indeed present on the
inner 10 kpc scale, based on the jet symmetry (Fig. 1).

We now highlight a few other striking properties of the present source 
that, when taken together, set it apart from nearly all the known FR~II 
radio galaxies.


(a) A super-massive   central black-hole is revealed by the very 
large stellar velocity dispersion $\sigma_{\ast}$ = 375$\pm$35 km s$^{-1}$ 
(Sect. 2.3).
Following \cite{Tremaine_2002}:

\begin{equation}
\log (M_{BH}/M_{\odot})=\alpha+\beta\log(\sigma_{\ast}/\sigma_{200}),
\end{equation}
\noindent
where $\sigma_{200} = 200$ km s$^{-1}$ and $\alpha=8.13\pm0.06, \, 
\beta=4.02\pm0.32$.
This gives for \cgcg ($\sigma_{\ast} = 353$ km s$^{-1}$) a black hole mass 
M$_{BH} = 1.32\pm0.93 \times 10^9$ M$_\odot$.

An independent estimate of M$_{BH}$ is obtained using the  
following relation
\citep{Marconi_Hunt03} :

\begin{equation}
\log (M_{BH}/M_{\odot})= a + b [log L_{K,bulge} - 10.9],
\end{equation}

\noindent
where $L_{K,bulge}$ is the bulge luminosity in {\it K}-band (2.17 $\mu$m)
in solar units $L_{K,\odot}$, and $a=8.21\pm0.07, \, b=1.13\pm0.12$.
Since there is no
indication of a disk component in CGCG~0049-033 \citep[SDSS;][]{SDSS},
its  {\it K}-band absolute magnitude ($M_{K}=-26.22\pm0.044$),
from the 2MASS data \citep{2mass}, gives its
bulge emission: $L_{K,bulge}= 6.5 \times 10^{11}
L_{K,\odot}$. The resulting M$_{BH} = (1.73\pm0.51) \times 10^9$ M$_\odot$,
is close to the above estimate using equation (1). The average
of the two,  M$_{BH} = (1.5\pm1) \times 10^9$ M$_\odot$,
falls  in  the league of most massive nuclear black-holes known
[e.g., the radio galaxy M87, with M$_{BH} = (2.4 \pm 0.7) \times 10^9$ $M_\odot$
\citep{Harms_1994}, and also the quasar 1745+624 \citep{Cheung_2006}].


(b) The observed strong linear polarization and its orientation imply a 
well organized transverse (toroidal) magnetic field for this giant FR~II 
jet, which, in principle, can result from compression  of initially tangled magnetic
fields in a succession of shocks \citep{Laing_1981}. However, the 
magnetic field in extended FR~II jets is typically found to be along 
the jet \citep{Bridle1994}. Conceivably, in  
kiloparsec scale jets
the dominant contribution to the magnetic field may arise from a dynamo in the
turbulent shear layers \citep[e.g.,][]{MS90,Urpin_2002}. The orthogonal magnetic 
field orientation
in the present jet is then also unusual (Sect. 4). 

(c) The observed core-to-lobe flux density ratio, $f_c = 1.67$, at 1.28 GHz is
an extreme value; for FR~II radio galaxies with matching radio lobe power to
CGCG~049-033, $f_c$ is typically only $\sim 0.1$ \citep{Zirbel_Baum95}. 
The large excess in $f_c$, by a factor 
$F \sim 17$, could be attributed to relativistic beaming, which can boost 
the core flux by a factor $\sim \delta^n$ (usually {\it n}$~\sim 2$ for 
compact flat-spectrum jets, and the Doppler factor is 
$\delta = [\Gamma_{j} (1-\beta_j \, 
cos \theta)]^{-1}$, where  $v_{j}= c \beta_j$ is the bulk speed of jet 
and $\Gamma_{j}$, the bulk Lorentz-factor). The corresponding 
viewing angle $\theta$ of the nuclear jet  
\citep[see][]{Giovannini_1994}, cos ($\theta$) = $[0.5 + (\sqrt{F} -1)/\beta_j]/\sqrt{F}$, 
is then quite small ($\theta \leqslant 28^{\circ}$ for $\beta_j \leqslant 1$). 
Such small angles  are jointly disfavored by  the projected
giant radio size, the apparent symmetry of the inner jets and the
identification with a galaxy \citep[see][]{Barthel_1989}. 

\section{Discussion}

The afore-mentioned peculiarities of CGCG~0049-033 should be viewed in
the context of its other striking abnormality, the compactness and
apparent one-sidedness of the radio lobe. 
Note that the counter-lobe
is undetected down to   brightness contrasts of $\sim$10 and $\sim$15
on the Effelsberg and NVSS maps, respectively (Fig.~2).
Indeed, the huge lobe asymmetry together with the symmetry of the 
inner jets stands in stark contrast to the pattern common for 
double radio sources \citep{Bridle1994,G-K-2005}.
We now discuss some possible scenarios for the lobe's unusual
properties.


\subsection{The Lobe Asymmetry: Delayed Lobe Formation?}

The appearance of the radio lobe on just one side of the nucleus is reminiscent 
of the quasar 3C~273, well known for its unique one-sided radio morphology 
\citep[e.g.,][]{Conway_Davis1993}, which may even be an artefact of 
the light travel time effect arising from a combination of 
small viewing angle ($\theta \sim$ $10^{\circ}$ - $20^{\circ}$) and 
the exceptionally large speed of the 
jet's head ($V_h$) \citep[e.g.,][]{KGK86,Stawarz2004}.

In CGCG~0049-033, the radio lobe resulting from  ``backflow'' of  
synchrotron plasma from the hotspot, is remarkably compact and well bounded 
even on the side facing the nucleus. Thus, no lobe is seen around the initial 
$\sim 80\%$ of the jet's length, giving the source a ``pendulum''-like 
appearance (Fig. 1). Moreover, since the non detection of the lobe of the 
counter-jet would require it to be intrinsically underluminous compared
to the main lobe by an extreme factor ($\gtrsim$ 20; Sect.~2), it seems
more likely that the northern lobe is hidden 
due to the light travel-time effects.
Thus, a plausible scenario is that the lobe formation in this source began 
only after the jet had grown to $\sim 80\%$ (i.e., $\sim 350$ kpc in 
projection) of its presently observed length ($\sim 440$ kpc in projection),
possibly due to an unusually prolonged phase of the jet's ballistic 
propagation. 
The observed radio extent  $\sim$100 kpc 
for the counter-jet is therefore only a lower limit, because 
its farthest part  could  be Doppler dimmed below the 
detection threshold. But, in any case, the outer edge of the source on the 
counter-jet side is now being monitored at most at a stage just prior to the 
onset of lobe formation (otherwise, that lobe too would have 
been detected). This assumes  intrinsic symmetry, which then also 
implies an upper limit of 350 kpc to the projected extent of the source 
on the counter-jet side (see above). 
The two limiting values of 100  and 350 kpc  correspond to a
ratio $R$ between 4.4 and 1.26, of the apparent lengths of the approaching
and receding jets, where
$R =  \left[ 1 + \left( {V_{h} \over c}\right) cos\theta \right] \slash
\left[ 1 - \left( {V_{h} \over c}\right) cos\theta \right]$ \citep{Stawarz2004}.

Therefore, $V_h > \,$c/3, 
even if
$\theta$ = 69$^{\circ}$,
the median value for FR~II radio sources in
the unified scheme
\citep{Barthel_1989}.  This is an order-of-magnitude
larger than the typical expansion speed of large radio
sources \citep[e.g.,][]{Barai_Wiita_2006} but similar
to the speed ($V_{h} \sim 0.85c$)
deduced by \cite{Stawarz2004} for the extremely asymmetric quasar 3C~273.
Stawarz has proposed
that the jet of 3C~273 is propagating through an ultra-rarified medium of a
fossil radio lobe (i.e., cavity) created during a previous
episode of nuclear activity.
Note that the delayed lobe formation scenario can ``hide'' the far-side 
lobe without requiring it to recede at a {\it bulk} relativistic speed.

\subsection{The Lobe's Compactness}

The lobe's striking compactness could be due to a 
dynamically important (toroidal) magnetic field associated with the jet, 
or a well ordered ambient magnetic field. The latter, too, could assist in 
jet collimation and formation of a compact radio lobe, as found in the 3-D 
magneto-hydrodynamical (MHD) simulations of relativistic jets propagating 
through a strong magnetic field aligned with 
the jet \citep{Nishikawa1997}. In the afore-mentioned scenario of 
jet propagation through a ghost cavity, 
such an aligned ambient magnetic 
field would indeed be expected for the jet \citep[see, e.g.,][]{FG2007}.

The detection of a highly organized transverse (toroidal) magnetic field
in the present jet is exceptional not only for the  largest physical 
scale found thus far, but also for being opposite to the usual pattern 
exhibited by FR~II jets (Sect. 3). This underscores the importance of this
mega-jet as a rare example where a dynamically important magnetic field  
is traceable up to  several hundred kiloparsecs from the nucleus. 
Plausibly, the light-travel time and peculiar magnetic configuration internal
(or external) to the jet are two rare effects simultaneously at work, 
leading to its extraordinary nature.
Finally, the sequence of radio knots 
observed at a quasi-regular spacing of $\sim 40$ kpc over most of
its length (Fig. 1) as well as  its excellent collimation make this mega-jet 
a prime target for probing jet confinement mechanisms. 
A deeper radio imaging of this jet could be used to test viability of
the theory of jet stabilization upto megaparsec scales, 
via a spine-sheath type flow \citep{Hardee2007}, or  the jet collimation 
by a surrounding high pressure ambient medium,
leading to a  sequence of reconfinement
shocks \citep{Komissarov_Falle_1998}.

\acknowledgments
We thank the operations teams of GMRT (NCRA.TIFR), Effelsberg (MPIfR), and  
Giravali (IUCAA) observatories. We  also thank 
Paul Wiita and Rainer Beck for  useful discussions.

\clearpage
\begin{figure}
\includegraphics[scale=1.5, clip=true]{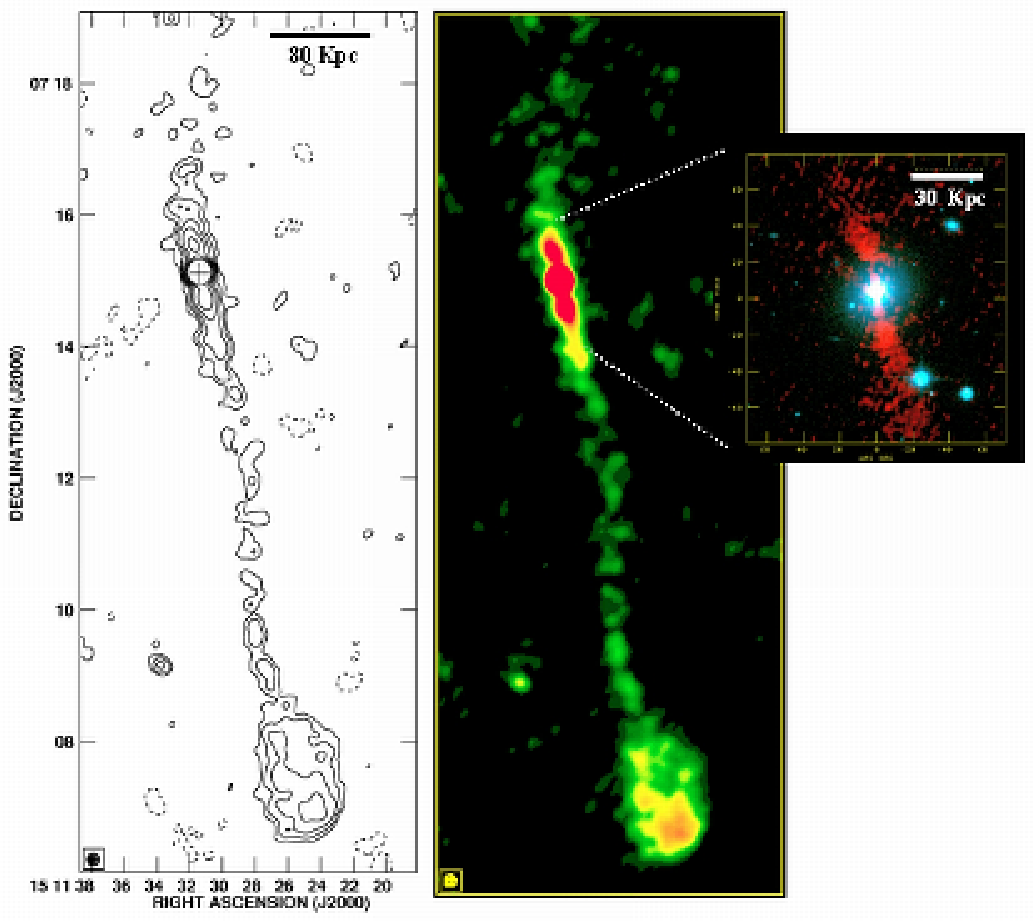}
\vskip 1cm
\caption{GMRT maps at 1.28~GHz, showing
{\it (left)}: Intensity contours: -0.18,0.18, 0.36, 0.72, 1.44, 3 and 
6 mJy/beam; rms
noise: $\sim60$ $\mu$Jy/beam, and the pseudo-color image {\it (center)},
both with a 11\arcsec\,  beam. The details of the inner $\sim$ 2\arcmin\,
region {\it (inset)} are visible in the 3\arcsec\, resolution 1.28~GHz
 GMRT  image (shown in red)
 overlaid on the  optical {\it r}-band SDSS image (shown in blue).}
\label{fig_GMRT}
\end{figure}
\newpage 
\begin{figure}
\includegraphics[angle=0,scale=0.33,clip=true]{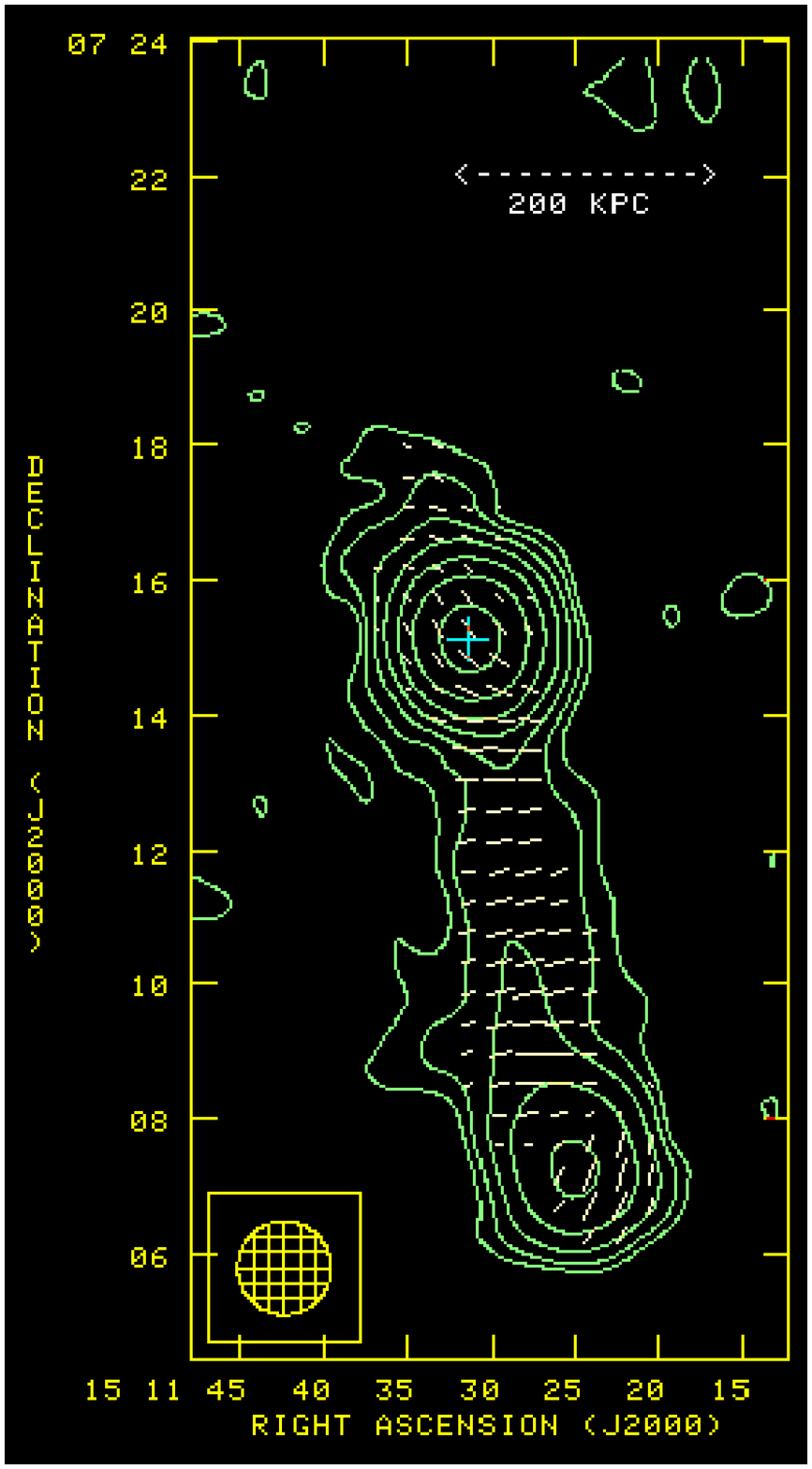}
\includegraphics[angle=0,scale=0.33,clip=true]{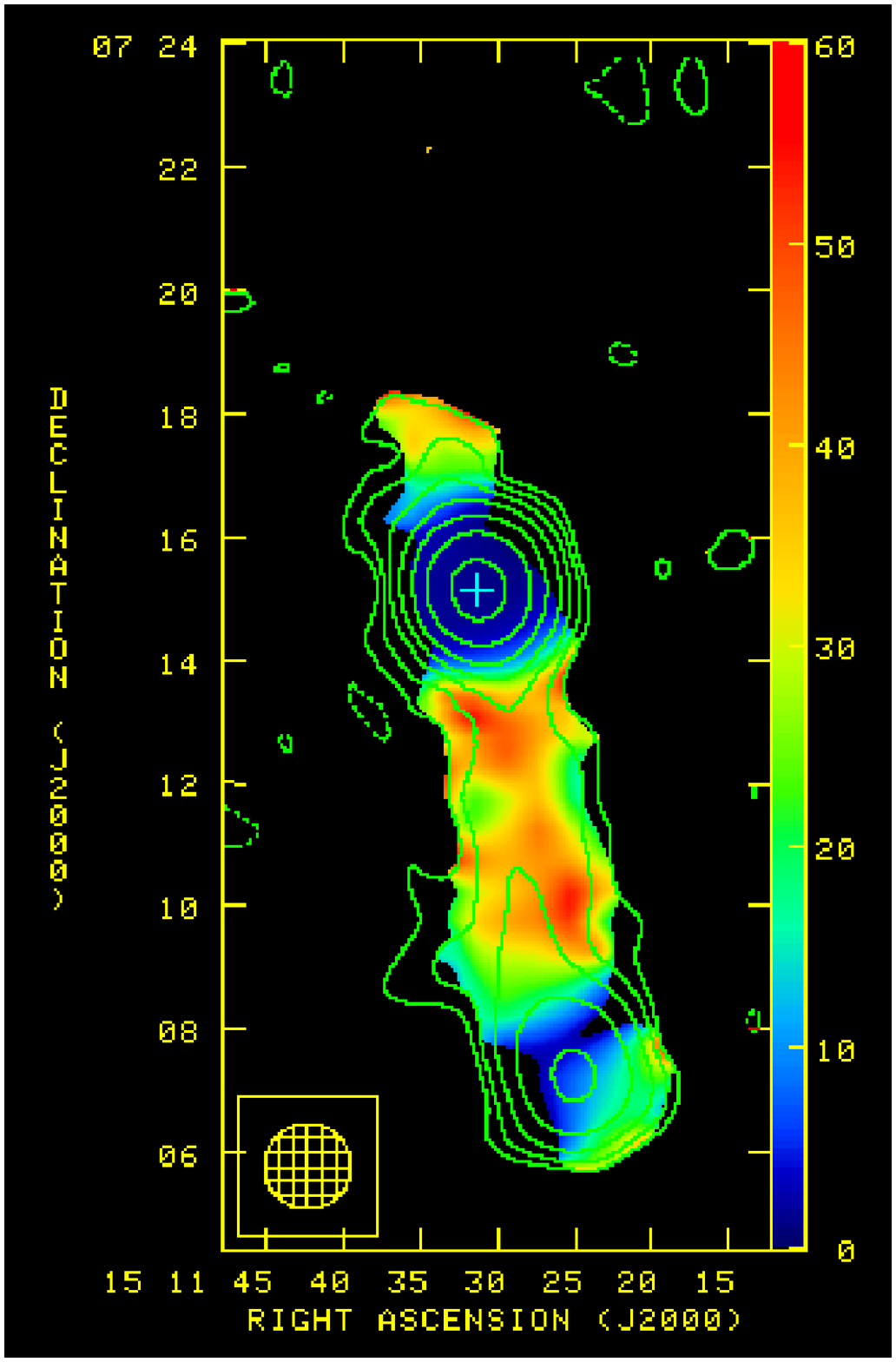}
\includegraphics[angle=0,scale=0.34,clip=true]{f2c.ps}
\includegraphics[scale=0.36,clip=true]{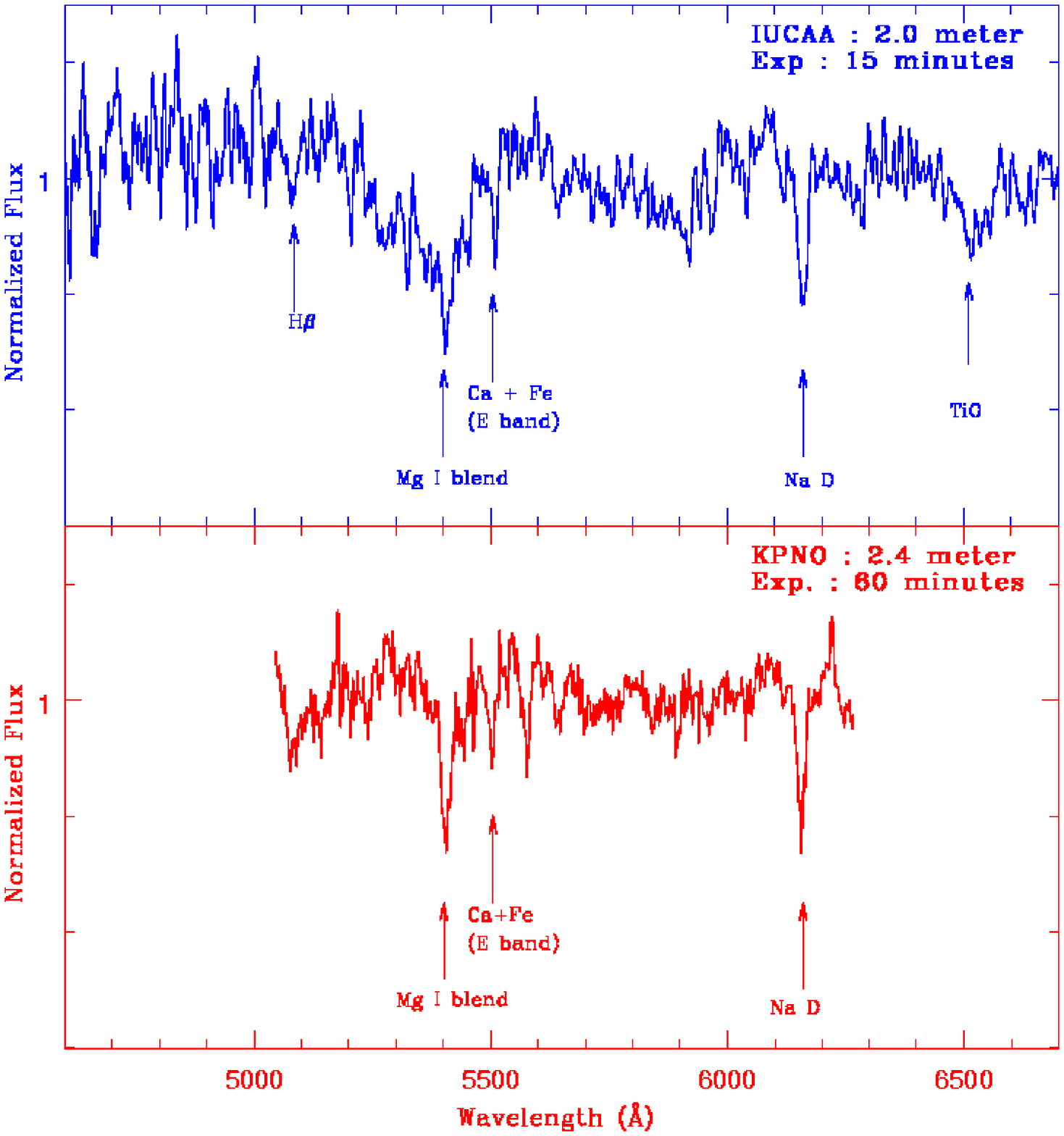}
\caption{{\it left}: Effelsberg  8.35 GHz  observations of \cgcg
showing, total intensity map
[contours: -0.75, 0.75, 1.5, 3, 6, 12,
24 and 48 mJy/beam] along with  rotation-measure (RM) corrected
magnetic field vectors (lines), having lengths
proportional to the local intensity of linearly polarized flux
(scale: 1\arcsec\, $=$  47.6 $\mu$Jy/beam). {\it middle}:
Percentage linear
polarization at 8.35 GHz shown
in pseudo-color, overlaid on the total
intensity contours. The HPBW  is 84\arcsec\, shown inside the box.
{\it right}: The 1.4 GHz NVSS contour map is shown 
(contours: -1.25, 
1.25, 2.5,
5, 10, 20, 40 and 80 mJy/beam), along with the electric field vectors of linear
polarization (scale: 1\arcsec\, $=$  0.13 mJy/beam).
The HPBW  is 45\arcsec\, shown inside a box.
The plus sign marks the radio core at the nucleus of
the elliptical galaxy CGCG~049-033 [R.A.= $15^{h} 11^{m} 31.38^{s}$,
Dec.=$07^{o} 15^{'} 7.11^{"}$ (J2000)].
{\it bottom}:
Optical spectra of \cgcg taken with IFOSC on the 2-m telescope of IUCAA
and with the 2.4-m Hiltner/KPNO telescope [for  details, Sect.~2.3].}
\label{fig_Effelsberg}
\end{figure}

\clearpage

\begin{thebibliography}{}

\bibitem[Barai \& Wiita 2006]{Barai_Wiita_2006} Barai, P., Wiita, P.J. 2006, 
\mnras, 372, 381

\bibitem[Barthel 1989]{Barthel_1989} Barthel, P.D. 1989, \apj,  336, 606


\bibitem[Bridle et al. 1994]{Bridle1994} Bridle, A.H.,  et al. 1994, \aj, 108, 766

\bibitem[Cheung et al. 2006]{Cheung_2006} Cheung, C.C., Stawarz, L., Siemiginowska,
A. 2006, \apj, 650, 679


\bibitem[Condon et al. 1998]{condon1998} Condon, J.J, et al. 1998, \aj, 115, 1693

\bibitem[Conway et al. 1993]{Conway_Davis1993} Conway, R.G., Garrington, S.T.,
Perley, R.A., Biretta, J.A. 1993, A\&A, 267, 347




\bibitem[Emerson \& Gr\"{a}ve 1988]{Emerson1988} Emerson, D.T., \& Gr\"{a}ve, R. 1988, A\&A, 190, 353 

\bibitem[Fanaroff \& Riley 1974]{FR74} Fanaroff, B.L., Riley, J.M 1974, \mnras, 167, 31p

\bibitem[Feretti \& Giovannini 2007]{FG2007} Feretti, L., Giovannini, G. 2007, astro-ph/0703494 (review)


\bibitem[Giovannini et al. 1994]{Giovannini_1994} Giovannini, G.,  et al. 1994, \apj,  435 ,116



\bibitem[Gopal-Krishna \& Wiita 2005]{G-K-2005} Gopal-Krishna, \&
Wiita, P.J., 2005,
{in `21st Century Astrophysics',}{ed. S. K. Saha
\& V. K. Rastogi, Anita Publications, New Delhi, page 108 (astro-ph/0409761)}


\bibitem[Hardee et al. 2007]{Hardee2007}
Hardee, P., Mizuno, Y., \& Nishikawa, K.-I. 2007, astro-ph/0706.1916

\bibitem[Harms et al. 1994]{Harms_1994} Harms, R.J.,  et al. 1994, \apj,  435, L35




\bibitem[Komissarov \& Falle 1998]{Komissarov_Falle_1998} Komissarov, S.S., Falle, S.A.E.G. 1998, \mnras, 297, 1087



\bibitem[Kundt \& Gopal-Krishna 1986]{KGK86} Kundt, W., \& Gopal-Krishna. 1986,
Jou. Astron. Astroph., 7, 225


\bibitem[Laing 1981]{Laing_1981} Laing, R.A. 1981, \apj,  248, 87

\bibitem[Liu \& Zhang 2002]{Liu_Zhang_2002} Liu, F.K., Zhang, Y.H 2002, A\&A, 381, 757

\bibitem[Marconi \& Hunt 2003]{Marconi_Hunt03} Marconi, A., Hunt, L.K. 2003, \apj, 589, L21




\bibitem[Matthews \& Scheuer 1990]{MS90}
Matthews, A., Scheuer, P.A.G. 1990, \mnras, 242, 623


\bibitem[Nishikawa et al. 1997]{Nishikawa1997} Nishikawa, K.-I., et al. 1997, \apj, 483, L45

\bibitem[Simard-Normandin et~al. 1981]{S-N1981} Simard-Normandin, M., Kronberg, 
 P.P, \& Button, S. 1981, \apjs, 45, 97 

\bibitem[Skrutskie et al. 2006]{2mass} Skrutskie, M.F., et al. 2006, \aj, 131, 1163


\bibitem[Stawarz 2004]{Stawarz2004} Stawarz, L. 2004, \apj, 613, 119


\bibitem[Swarup et al. 1991]{Swarup1991} Swarup, G., et al. 1991, Current Science, 60, 95

\bibitem[Tremaine et al. 2002]{Tremaine_2002} Tremaine, S., et al. 2002, \apj, 574, 740

\bibitem[Urpin 2002]{Urpin_2002} Urpin, V. 2002, A\&A, 385, 14

\bibitem[Wegner et al. 1999]{wegner1999} Wegner, G., et al. 1999, \mnras, 305, 259

\bibitem[Willis et al. 1982]{Willis1982} Willis, A.G., Strom, R.G., 
Perley, R.A., 
\& Bridle, A.H. 1982, in IAU Symp. 97,
Extragalactic Radio Sources, 
ed. D.S. Heeschen \& C.M. Wade (Dordrecth: Reidel), 141

\bibitem[York et al. 2000]{SDSS} York, D.G., et al. 2002, \aj,  120, 1579

\bibitem[Zirbel \& Baum 1995]{Zirbel_Baum95} Zirbel, E.L., Baum, S.A. 1995, \apj, 448, 521

\end{thebibliography}
\end{document}